\documentclass[journal,a4paper]{IEEEtran}
\usepackage[pdftex]{graphicx}
\usepackage{wasysym} 
\usepackage{color}

\newcommand{\um}{\mathrm{\mu m}}
\newcommand{\Tc}{T_\mathrm{c}}
\newcommand{\mocm}{$\mu\Omega$ cm}

\begin{document}
\bstctlcite{IEEEexample:BSTcontrol}

\title{Superconducting NbTiN Thin Films with Highly Uniform Properties over a \diameter 100 mm Wafer}

\author{D.J.~Thoen,
        B.G.C.~Bos,
         E.A.F. Haalebos,
        T.M. Klapwijk,
        J.J.A. Baselmans,
        and~A.~Endo
\thanks{D.J. Thoen, J.J.A. Baselmans, and A. Endo are with the Department of Electrical Engineering, Faculty of Mathematics and Computer Science (EEMCS), Delft University of Technology, Mekelweg 4, 2628 CD Delft, The Netherlands. e-mail: (see http://terahertz.tudelft.nl/).}
\thanks{D.J. Thoen, B.G.C. Bos, T.M. Klapwijk, and A. Endo are also with the Kavli Institute of NanoScience, Faculty of Applied Sciences, Delft University of Technology, Lorentzweg 1, 2628 CJ Delft, The Netherlands.}
\thanks{E.A.F. Haalebos and J.J.A. Baselmans are with the Netherlands Institute for Space Research (SRON), Sorbonnelaan 2, 3584 CA Utrecht, The Netherlands.}
\thanks{T.M. Klapwijk is also with the Physics Department, Moscow State Pedagogical University, 119991 Moscow, Russia.}
}

\maketitle

\begin{abstract}

Uniformity in thickness and electronic properties of superconducting niobium titanium nitride (NbTiN) thin films is a critical issue for upscaling superconducting electronics, such as microwave kinetic inductance detectors for submillimeter wave astronomy. In this article we make an experimental comparison between the uniformity of NbTiN thin films produced by two DC magnetron sputtering systems with vastly different target sizes: the Nordiko 2000 equipped with a circular \diameter100 mm target, and the Evatec LLS801 with a rectangular target of 127 mm $\times$ 444.5 mm. In addition to the films deposited staticly in both systems, we have also deposited films in the LLS801 while shuttling the substrate in front of the target, with the aim of further enhancing the uniformity. Among these three setups, the LLS801 system with substrate shuttling has yielded the highest uniformity in film thickness ($\pm $2\%), effective resistivity (decreasing by 5\% from center to edge), and superconducting critical temperature ($T_{\mathrm{c}}$ = 15.0 K - 15.3 K) over a \diameter100 mm wafer. 
However, the shuttling appears to increase the resistivity by almost a factor of 2 compared to static deposition. Surface SEM inspections suggest that the shuttling could have induced a different mode of microstructural film growth. 

\end{abstract}


\section{Introduction}

Superconducting niobium titanium nitride (NbTiN) thin films are used for highly demanding circuits that operate in the frequency range of 1 GHz - 1000 GHz.
Having a gap frequency of $F_{\mathrm{gap}}\sim$1100 GHz, 
NbTiN is being used in transmission lines for astronomical instruments that operate at frequencies above the gap frequency of Nb ($F_{\mathrm{gap}}\sim$700 GHz) \cite{deGraauwHIFI,Jackson2006}.
Aside from the high $F_{\mathrm{gap}}$, NbTiN is known to exhibit little microwave phase noise \cite{Barends2010noise} and microwave loss \cite{Barends2010Loss,Bruno2015}, making it a good material for photodetectors \cite{Janssen2013eff} and circuit quantum electrodynamic experiments \cite{vanWoerkom:2015jt}.
Furthermore, NbTiN is also being used as the material for 
narrow band filters \cite{Endo2013}, and microwave parametric amplifiers \cite{Eom2012}.
In the above-mentioned applications, the typical size of each chip has been on the order of 0.1 mm - 10 mm.
However, recent applications of NbTiN have a rapidly growing degree of on-chip multiplexing, demanding the chip size to grow to 100 mm and beyond. For example, upcoming submillimeter astronomical instruments such as A-MKID \cite{Baryshev2014} and DESHIMA \cite{Endo2012SPIE} demand $10^3$-$10^4$  of NbTiN/Al hybrid microwave kinetic inductance detectors (MKIDs) \cite{Janssen2013eff}, which fill the entire surface of one or more \diameter 100 mm wafers. 
Such large-scale devices have a dramatically higher demand in the uniformity of critical film properties over a large surface area, and often d.c. magnetron sputtering methods that have been designed for small devices cannot easily meet the requirements. 

\begin{figure}[!t]
\centering
\includegraphics[width=85mm]{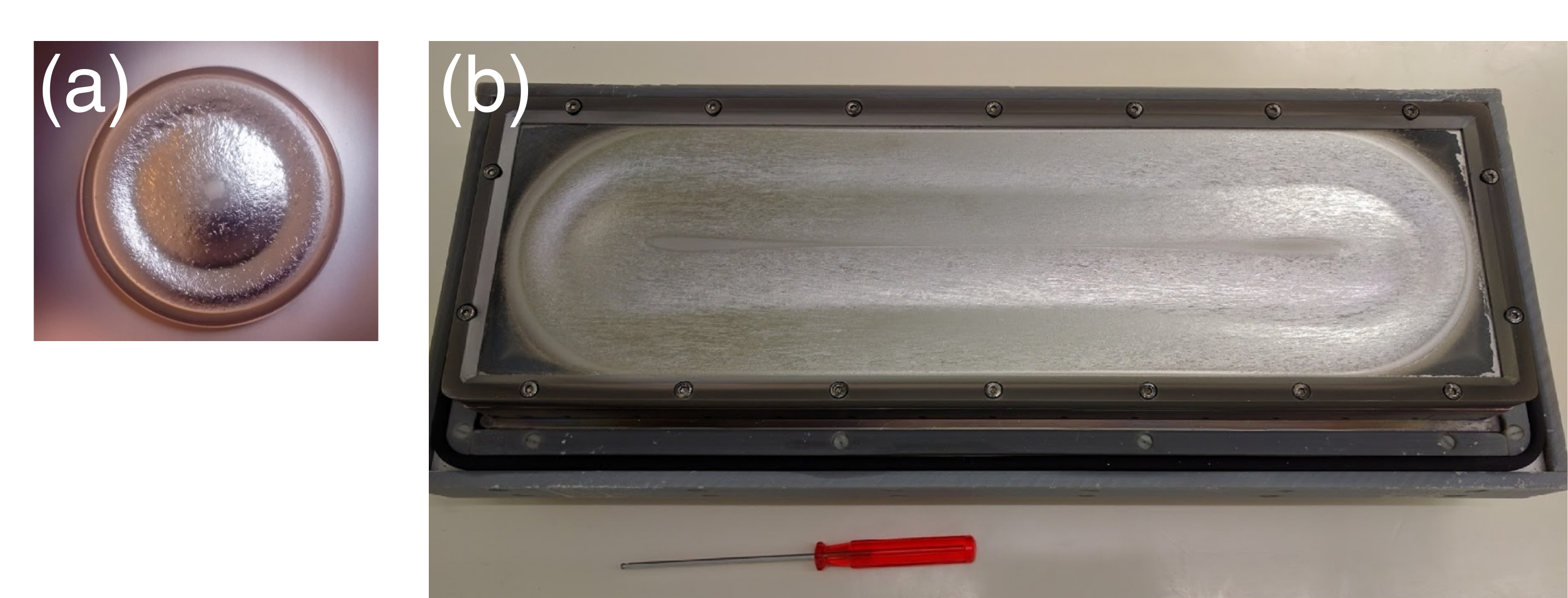}
\caption{The targets of the two instruments that were used, presented roughly in scale with each other. (a) Nordiko 2000 NbTi sputtering target, \diameter 100 mm, sputtering track $\sim$\diameter60 mm. (b) LLS801 NbTi sputtering target, 444.5 mm long, 127 mm wide.}
\label{Target}
\end{figure}

Here we investigate how the uniformity in thickness and physical properties of NbTiN films changes, when: (1) we adopt a sputtering target that is much larger than a \diameter 100 mm wafer (especially in one direction), and (2) shuttle the wafer under the target during deposition (Fig. $\ref{Target}$). 
Using a larger sputter system enables us to scale deposition parameters from the smaller system to allow for a relatively easy transition between the two deposition systems \cite{bos}. Other techniques to obtain a better homogeneity, such as confocal sputtering, result typically in a strong reduction in deposition speed which can have additional complications such as increased impurity concentrations and a different film growth. 

In this paper we will especially focus on the parameters that are relevant for large arrays of MKIDs.
In a typical MKID chip, $\sim$1000 MKIDs can be read out simultaneously with a bandwidth of 2 GHz \cite{vanRantwijk2016}, relying on the assumption that the resonance frequencies follow the design with a precision of no worse than $\sim$2 MHz. While the physical length of each resonator can be controlled with sufficient precision by standard lithographic techniques, it is challenging to keep the kinetic inductance $L_\mathrm{k}$ uniform enough over the large device area to prevent the resonance features from overlapping with one another. Because $L_\mathrm{k}$ is ultimately associated to the critical temperature $T_\mathrm{c}$, resistivity $\rho$, and film thickness $t$ (see Appendix), we will investigate the uniformity of these parameters.


\section{Experiment}
\label{experiment}

\subsection{Sputtering machines and targets}
We have used two sputtering machines in the experiments: a Nordiko 2000 system, a research-based sputtering machine from Nordiko Technical Services Ltd. and the LLS801, a refurbished and customized industry-based Evatec LLS801 sputtering system. The Nordiko is well explored for reactive NbTiN deposition \cite{Iosad2001} but suffers from a lack of uniformity as we will be presented in Section \ref{results}. The thickness uniformity of non-reactive deposited films (e.g., Al) of LLS801 is $\sim$2\%. Further details regarding these two sputter machines can be found in \cite{bos}. Operation of the reactive plasma in order to acquire high-quality NbTiN has been studied by \cite{bos}. 

The targets used in both systems are presented in Fig. $\ref{Target}$, with their dimensions. The targets are purchased at Thermacore Inc. while fabrication took place at their Materials Technology Division in Pittsburgh, PA, USA. The targets are made from the same batch of raw material, which has been double arc-melted. The casting was severed. One part is machined into multiple round targets while the other part was melted a third time, followed by machining the rectangular target. The targets have a composition of 81.9 wt.\% Nb and 18.1 wt.\% Ti, with a purity 99.95\%. Particle analysis shows that magnetic impurities have been reduced to 30 parts per million (PPM) for oxygen, and less than the ICP-AES measurement limit for iron $<$23.1 PPM, chromium $<$9.23 PPM and nickel $<$4.62 PPM.

In the LLS801, the substrates are mounted on the surface of a cylindrical drum of 670 mm in diameter. The target is situated outside of the drum, pointing normally at the substrate in static deposition mode ($\phi=0^\circ$). In shuttle deposition mode, the drum rotates back and forth in the direction of the short side of the target, over an angular range of $\phi = \pm$20$^\circ$ at a speed of 4.5$^\circ \ \mathrm{s}^{-1}$.

\subsection{Sample preparation and measurement}

In order to compare the effect of the size of the target and wafer shuttling on the uniformity, 
we have deposited NbTiN films on three wafers in the following manner: 1) static deposition using the Nordiko 2000, 2) static deposition using the LLS801, 3) shuttled deposition using the LLS801. 
For all three methods, the plasma conditions had been optimized to make $T_\mathrm{c}$ at the center of the wafer as high as possible \cite{iossad, bos}. For the Nordiko 2000, this meant an Ar partial pressure of 0.5 Pa, Ar flow of 100 sccm, N$\mathrm{_2}$ flow of 7.5 sccm and an applied power of 440 W. Similarly for the LLS801, Ar partial pressure 0.7 Pa, Ar flow 400 sccm, N$\mathrm{_2}$ flow 84.7 sccm and 5.0 kW power. 

All films were deposited on sapphire wafers (Kyocera, C-plane, \diameter100 mm, thickness 0.36 mm, double side polished), which were cleaned with phosphoric acid. 
After film deposition, electron beam lithography and reactive ion etching are used to pattern a 5 mm $\times$ 5 mm grid of test structures. A single pattern contains one wire with a width of $W$ = 150 $\mu$m and length of $L$ = 450 $\mu$m, connected to probing pads for four-point probe measurements. Using this pattern, we conducted position depended measurements of $t$, square resistance $R_{\Box}$ and $\Tc$, from which we calculate the effective resistivity $\rho = R_{\Box}t$. 
The obtained $t$, $R_{\Box}$ and $\Tc$ are used to calculate the magnetic penetration depth $\lambda_\mathrm{m}$  as well as the change in frequency due to variations in film properties (see Appendix).

\begin{figure*}[!ht]
\centering
\includegraphics[width=0.93\textwidth]{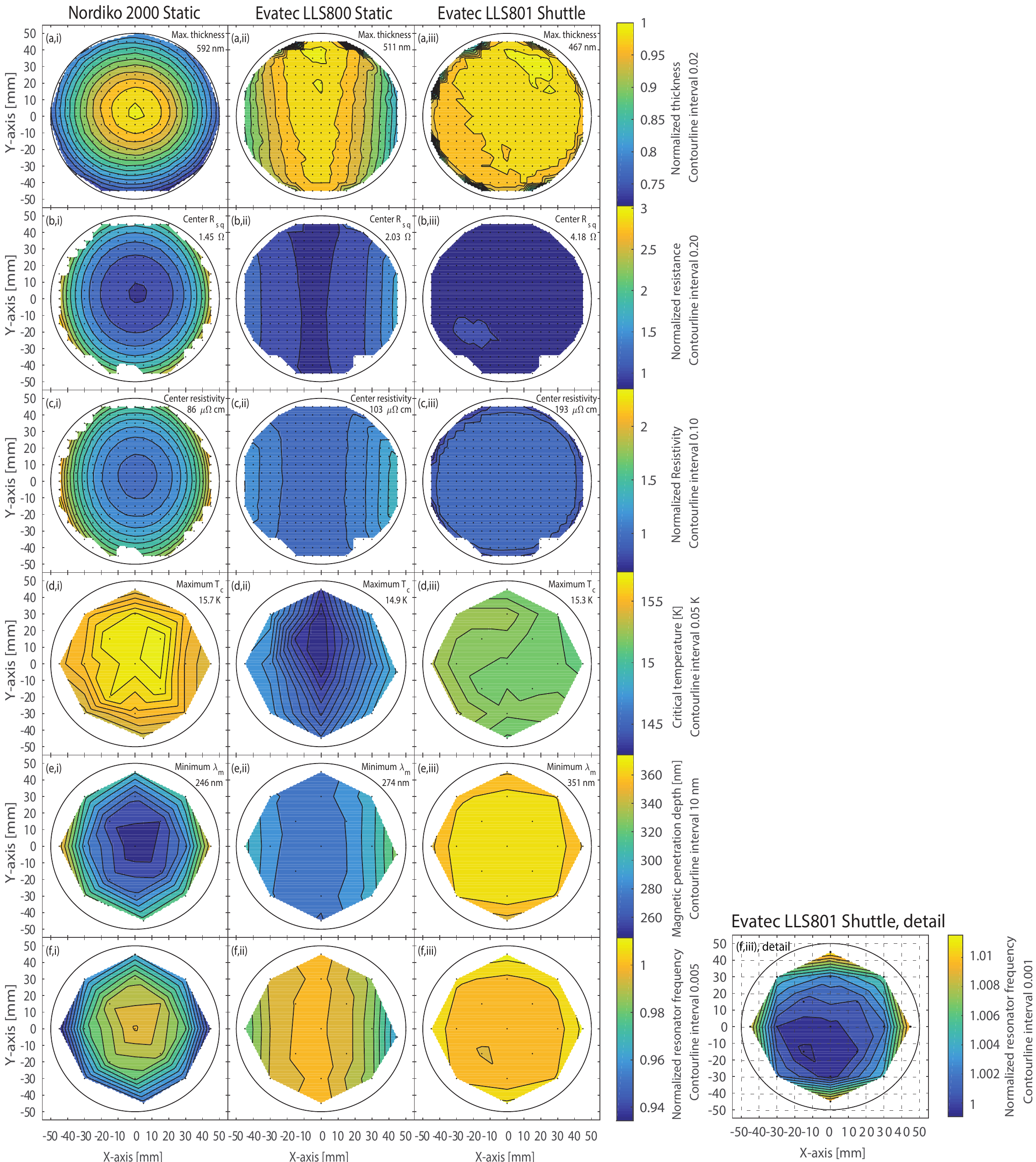}
\caption{Contour/intensity plots of measured and calculated physical parameters over the \diameter 100 mm area of the wafers. (a): Normalized thickness with respect to the maximum thickness, the interval of the contour lines is 0.02. 
(b): Normalized square resistance with respect to the center position. The interval of the contour lines is 0.20. 
(c): Normalized room temperature resistivity with respect to the center position. The interval of the contour lines is 0.10.
(d): Critical temperature. The interval of the contour lines is 0.05 K.
(e): Magnetic penetration depth. The interval of the contour lines are 10 nm.
(f): Normalized resonator frequency with respect to the center position as calculated by Eq. \ref{deltaF}. The interval of the contour lines is 0.005.
(i): Nordiko 2000, (ii): LLS801 static, (iii): LLS801 shuttle. (a) - (d): Each black dot represents an individual measurement point, while for (e) and (f) it represents a point of calculation. Fig. \ref{Uniformity} (f,iii) detail: Detail image of the normalized resonator frequency data of the LLS801 shuttled wafer. The variation in $\delta F$ becomes visible at an interval level 0.1\%. 
}
\label{Uniformity}
\end{figure*}

\section{Results and discussion}
\label{results}

If Fig. \ref{Uniformity}, we present the measured thickness $t$ (row a), $R_{\Box}$ (row b) and $\Tc$ (row d) for the 3 wafers deposited by means of static deposition in the Nordiko 2000 (column i), static deposition in the LLS801 (column ii), and shuttle deposition in the LLS801 (column iii). From these values we have calculated the effective resistivity $\rho$ (row c), the magnetic penetration depth $\lambda_\mathrm{m}$ (row e), and the shift in resonance frequency of a typical microwave resonator (row f), as explained in Section \ref{experiment}.

The effect of enlarging the target from a \diameter 100 mm circle (Nordiko 2000) to a 127 mm $\times$ 444.5 mm rectangle (LLS801) can be seen by comparing columns (i) and (ii) in Fig. \ref{Uniformity}. Regarding $t$ and $R_{\Box}$, the film from the Nordiko 2000 shows clear dome-shaped, concentric circular patterns, whilst the film from the LLS801 shows dike-shaped patterns evolving from the center in the $x$ direction but keeping a relatively constant value in the $y$ direction, which are aligned to the short and long sides of the rectangular target, respectively. In the Nordiko 2000, the $t$ and $R_{\Box}$ values change rapidly outside a diameter of $\sim$50 mm, which is reasonable because the system was originally intended for \diameter 50 mm wafers. In the case of the LLS801, the differences in maximum to minimum $t$ and $R_{\Box}$ are significantly smaller compared to the Nordiko 2000, showing the advantage of a larger target and its larger erosion track. 

However, for the $T_\mathrm{c}$ distribution, the advantage of the larger target of the LLS801 is negligible. Whilst for the Nordiko 2000 the $T_\mathrm{c}$ exhibits a concentric pattern that resembles those for the $t$ and $\rho$, with the $T_\mathrm{c}$ decreasing from the center to the edge, the $T_\mathrm{c}$ of the film from the LLS801 shows a striking {\it minimum} at $x\sim 0$, even though the resistivity is lowest there. This is possibly caused by not exactly perfect sputter conditions that result in a nitrogen deficit in the wafer center.


This remaining non-uniformity in the films from the larger target of the LLS801 is greatly improved by the shuttling motion of the wafer, as can be seen by comparing columns (ii) and (iii) of Fig. \ref{Uniformity}. The non-uniformity in the $x$ direction that was present in the static deposition has been completely removed by shuttling. Over the central \diameter80 mm where the data is most complete and reliable, the thickness varies by only $\pm$2\%, the resistivity drops by 4\% from the center to the edge, and the $T_\mathrm{c}$ is 15.2 K within $\pm$1\%. This naturally results in a very uniform distribution for the resistivity and magnetic penetration depths, as can be seen in rows (c) and (e). Finally, we combine these values to calculate a resonance frequency shift for a typical microwave resonator at 5 GHz with a line width of $S$ = 3 $\um$ and slot width of $W$ = 2 $\um$. As shown in row (f), we see that for the shuttled deposition the frequency decrement is within 1\% over the whole wafer, while for an inner disk of 50 mm diameter this variation is limited to 0.2\%. In figure \ref{Uniformity} (f, iii detail) we show a close-up of the deviation in $\delta F$ with contour intervals of 0.1\%.  


While the uniformity of films produced by shuttle deposition is excellent, the high critical temperature of 15.3 K in combination with a relatively large specific resistivity of 200 \mocm$ $ is unexpected (about a factor of 2 higher compared to static deposition as can be seen in row (c) of Fig. \ref{Uniformity}). It is known that all thin films of normal metals \cite{mooij} show a correlation between resistivity and the temperature coefficient, changing sign at a resistivity of a few hundred \mocm, known as the Mooij-correlation. This empirical observation is in subsequent work connected to strong localization due to quantum interference of scattered trajectories and enhanced electron-electron interaction \cite{imry} with at strong disorder the tendency to become an insulator at low temperatures. This approach to an insulator competes in superconducting materials with the zero-resistance superconducting state. This competition has been experimentally documented in materials such as TiN \cite{sacepe}, NbN \cite{kamlapure} and effects of this behavior have also been found in NbTiN \cite{driessen}. The key point is that a high resistivity irrespective of its source weakens the superconductivity, as shown in a lower $\Tc$ and a gradual appearance of a spatially fluctuating order parameter. Therefore a resistivity of 200 \mocm$ $ and a high $\Tc$ are expected to be mutually incompatible. 

\begin{figure}[t]
\centering
\includegraphics[width=85 mm]{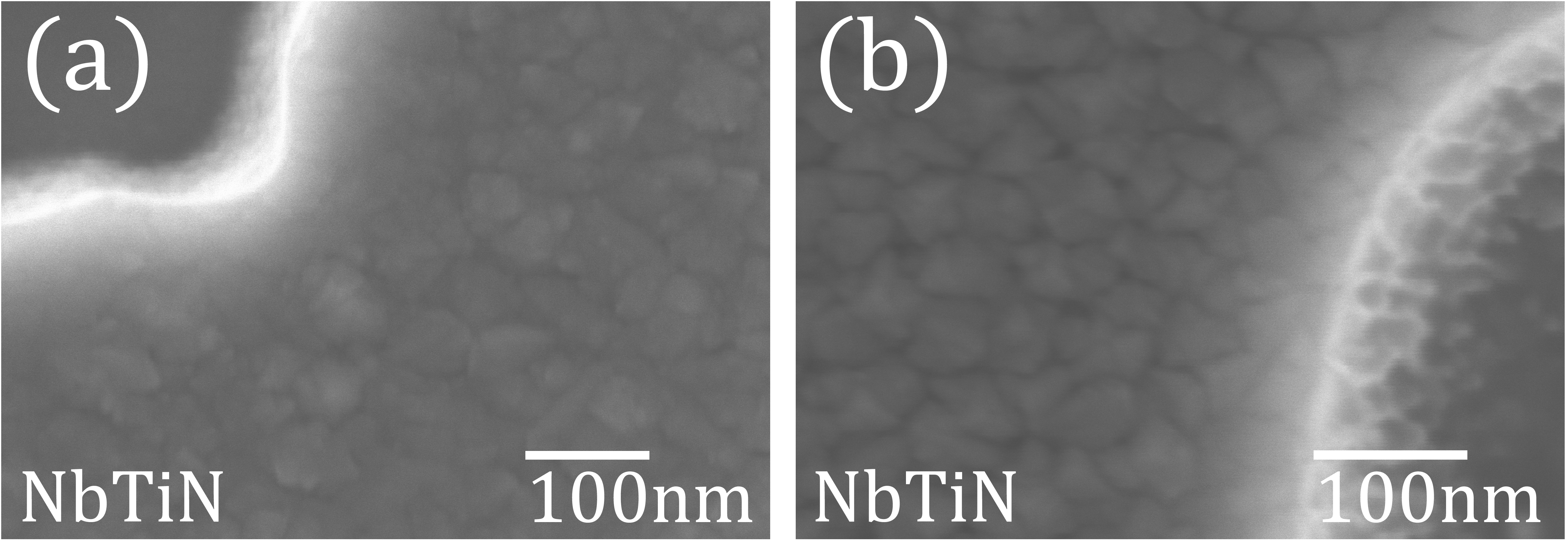}
\caption{Scanning electron micrographs of the surface of the NbTiN films produced in the LLS801, using static (left) and shuttle (right) deposition.}
\label{SEM}
\end{figure}

We have compared the different films by a scanning electron microscope (SEM) to compare the surfaces of the three deposited films. The micrographs are presented in Fig. \ref{SEM}. While the staticly deposited film shows a dense packing of columns (Thornton zone T) with horizontal sizes ranging from 20 nm to 60 nm, the shuttle-deposited film appear to exhibit deep voids between columns (Thornton zone 1), with typical sizes of $\sim$50 nm. This suggests that the higher resistivity of the shuttle-deposited film could be a result of a different mode of columnar growth of the NbTiN film \cite{iossad}, possibly caused by the reduced average energy flux induced by the shuttling because the wafer is not always exposed to the flux of sputtered material. The change in microstructural growth from zone 1 to zone T \cite{petrov} is further verified by the intrinsic film stress \cite{muller}, which is compressive at -418 MPa for the staticly deposited film and tensile at +24 MPa for the shuttled film. So the enhanced resistivity could be caused by this change in film growth. However, it leaves the question of the surprisingly high $\Tc$. 

An alternative explanation is based on the nature of the shuttled process. The deposition and growth of the thin film is delicate balance between the arrival rate of atoms, nitrogen and argon. Unavoidably, in the shuttled process these conditions change in an oscillatory manner, which makes it likely that the film composition over the thickness varies periodically during the growth of the film. During one deposition, the substrate passes the target about 40 times. Each time, the substrate receives a fresh layer of high $\Tc$ NbTiN, around 10 nm in thickness, which is then covered by a thin layer of a less optimal deposition, when the substrate leaves the deposition flux. Therefore, we assume that in our shuttled films good superconducting material with a low resistivity and a high Tc is interleaved with poorer quality material with a high resistivity and a low $\Tc$. For a $\Tc$ measurement we use the resistive transition, which measures selectively the layers with the highest critical temperature. For the resistivity determination we assume a uniform film whereas in reality it consists of a parallel circuit, with the lowest resistivity dominating, which may account for one half of the total film thickness and therefore effectively a resistivity of a factor of 2 lower. Further experiments are needed to test such an interpretation, which is important for further optimization towards large-area uniform films.

\section{Conclusion}

Enlarging the sputtering target and shuttling the substrate in front of it have both proven to be effective methods for improving the uniformity in thickness and electronic properties of superconducting NbTiN thin films. By combining the two, we have obtained a \diameter 100 mm circular film with $T_\mathrm{c}$ = 15.2 K $\pm$ 1\%, and the variations in thickness and other physical properties are also kept to within 1-2\%. This type of film is very suited for MKID applications.


\section*{Acknowledgment}
The authors would like to thank Marcel Bruijn and Vignesh Murugesan, members of the SRON clean room staff, for their support in the fabrication and processing at SRON. This research was supported by the NWO (Netherlands Organisation for Scientific Research) through the Medium Investment grant (614.061.611). AE was supported by the NWO Vidi grant (639.042.423). TMK was supported by the Ministry of Science and Education of Russia under Contract No. 14.B25.31.0007 and by the European Research Council Advanced Grant No. 339306 (METIQUM). 

\appendix
\label{appendix}

The magnetic penetration depth $\lambda_\mathrm{m}$ of an extremely dirty superconductor like NbTiN for $T\to 0$ is approximately \cite{Bartolf:2016ka}
\begin{equation}
	\lambda_\mathrm{m} = \sqrt{\frac{\hbar\rho}{\pi \mu_0 \Delta_0}}\sim 105 (\mathrm{nm})\times\sqrt{\frac{\rho\ (\mathrm{\mu \Omega\ cm})}{T_{\mathrm{c}}\ (\mathrm{K})}},
\end{equation}
where $\hbar$ is the Dirac constant, $\mu_0$ is the permeability of vacuum, and $\Delta_0\sim 1.764k_{\mathrm{B}}\Tc$ is the superconducting gap energy at 0 K ($k_{\mathrm{B}}$ is the Boltzmann constant). The kinetic inductance per unit length $L_\mathrm{k}$ of a CPW is given by
\begin{equation}
	L_\mathrm{k}=gL_\mathrm{s} = g\mu_0\lambda_\mathrm{m} \coth \Bigl( \frac{t}{\lambda_\mathrm{m}}\Bigr),
\end{equation}
where $L_\mathrm{s}$ is the surface inductance of the film, and $g(S, W, t)$ is the geometry factor for a CPW \cite{BarendsPhDthesis} ($g\sim 4 \times 10^5$ for the parameter sets in this work). 
For each position ($x$,$y$) on the wafer, we interpolate the measured $T_c$, $R_\Box$, and $t$ to calculate the resonance frequency shift of a CPW resonator according to 
 \begin{equation}
    \frac{F_0(0,0)+\delta F(x,y)}{F_0(0,0)}=\sqrt{\frac{L_k(0,0)+L_g}{L_k(x,y)+L_g}},
    \label{deltaF}
\end{equation}
where $L_g$ is the geometric inductance of the CPW \cite{BarendsPhDthesis}.



%

\bibliographystyle{IEEEtran}

\bibliography{Final}

\end{document}